\documentstyle[12pt]{article}
\font\tenrm=cmr10

\newcommand{\eq}{\begin{equation}}
\newcommand{\en}{\end{equation}}
\newcommand{\eqn}{\begin{eqnarray}}
\newcommand{\enn}{\end{eqnarray}}
\newcommand{\CR}{\nonumber \\}

\newcommand{\bG}{\bar{G}}
\newcommand{\BG}{{\bf g}}
\newcommand{\BH}{{\bf h}}
\newcommand{\cD}{{\cal D}}
\newcommand{\I}{{\rm i}}
\newcommand{\half}{{1\over2}}
\newcommand{\HBG}{\hat{{\bf g}}}
\newcommand{\pa}{\partial}
\newcommand{\str}{{\rm str}}
\newcommand{\tr}{{\rm tr}}
\newcommand{\mat}[2]{\left(\begin{array}{#1}#2\end{array}\right)}
\newcommand{\zw}[2]{{#2 \over (z-w)^{#1}}}
\newcommand{\A}{\alpha}

\newcommand{\D}{\delta}
\newcommand{\DE}{\Delta}

\newcommand{\vep}{\varepsilon}

\newcommand{\LM}{\Lambda}
\newcommand{\lm}{\lambda}
\newcommand{\vp}{\varphi}
\newcommand{\LB}{| \! [}
\newcommand{\RB}{] \! |}
\begin{document}
\renewcommand{\thefootnote}{\fnsymbol{footnote}}
\renewenvironment{thebibliography}[1]
  { \begin{list}{\arabic{enumi}.}
    {\usecounter{enumi} \setlength{\parsep}{0pt}
     \setlength{\itemsep}{3pt} \settowidth{\labelwidth}{#1.}
     \sloppy
    }}{\end{list}}
\parindent=1.5pc
\begin{titlepage}
\begin{flushright}
NBI-HE-92-77 \\
October 1992
\end{flushright}
\vspace{3cm}
\begin{center}{{\bf THE $W$ ALGEBRA STRUCTURE \\
OF $N=2$ $CP_{n}$ COSET MODELS}
\footnote{To appear in the proceedings of the International
Workshop on \lq\lq String Theory,
Quantum Gravity and the Unification of Fundamental Interactions"
Rome, September 21--26, 1992}
\\
\vglue 5pt
\vglue 1.0cm
{KATSUSHI ITO}\footnote{Address after March 1, 1993:
Institute of Physics, University of Tsukuba, Ibaraki 305, Japan}\\
\baselineskip=14pt
{\it Niels Bohr Institute,
     Blegdamsvej 17, DK-2100 Copenhagen {\O}, Denmark}\\
\vglue 0.8cm
{ABSTRACT}}
\end{center}
\vglue 0.3cm
{\rightskip=3pc
 \leftskip=3pc
 \tenrm\baselineskip=12pt
 \noindent
We discuss the $N=2$ super $W$ algebras from the hamiltonian
reduction of affine Lie superalgebras $A(n|n-1)^{(1)}$ and
$A(n|n)^{(1)}$. From the quantum hamiltonian reduction of
$A(n|n-1)^{(1)}$ we get the free field realization of $N=2$ $CP_{n}$
super coset models. In the case of the affine Lie superalgebras
$A(n|n)^{(1)}$, the corresponding conformal field theories do not have
 $N=2$ superconformal symmetry.
However we show that these models are twisted $N=2$ $CP_{n}$ models
and may be regarded as topological conformal field theories.
\vglue 0.8cm}
\end{titlepage}
\newpage
\renewcommand{\thefootnote}{\arabic{footnote}}
\setcounter{footnote}{0}
{\bf\noindent 1. Introduction}
\vglue 0.2cm
The $W$-algebra symmetry$^{1}$ in two dimensions plays a fundamental
role in various integrable systems such as generalized KdV hierarchies
and Toda field theories.
In conformal field theories the $W$ algebras provide an important class
of chiral algebras and have been studied extensively$^{2}$.
A systematic construction of the $W$-algebra associated with a simple Lie
algebra is given by the hamiltonian reduction of the affine Lie
algebra$^{3,2}$.
One of the recent developments in this direction is that superconformal
symmetries and their $W$-extensions are also obtained by considering
the hamiltonian reduction of a certain class of affine Lie
superalgebras$^{4}$.
In particular, the $W$-extension$^{5-11}$ of the $N=2$ superconformal
symmetry is an interesting subject in view of its relation to
topological field theories and the compactifications of superstrings.

In previous papers$^{5,6}$, the author has studied the quantum
hamiltonian reduction of the affine Lie superalgebras $A(n|n-1)^{(1)}$
and the Feigin-Fuchs representations of the $N=2$ super $W$ algebras,
which characterize the $N=2$ super $CP_{n}$ coset models constructed
by Kazama and Suzuki$^{12}$ (see also ref. 7).
Inami and Kanno$^{9}$ observed that the classical $N=2$ super $W$
algebra appears also in the $N=2$ super KdV hierarchies associated
with the affine Lie superalgebras $A(n|n)^{(1)}$.

The purpose of the present talk is to explain why these different Lie
superalgebras correspond to the same $N=2$ super $W$ algebra.
We shall discuss the classical and quantum hamiltonian
reduction of the affine Lie superalgebras $A(n|n-1)^{(1)}$
and $A(n|n)^{(1)}$.
We will show that the conformal field theories associated with the affine
Lie superalgebras $A(n|n)^{(1)}$, are the twisted $N=2$ $CP_{n}$ models.

\vglue 0.6cm
{\bf\noindent 2. Lie superalgebras and their hamiltonian reductions}
\vglue 0.4cm
{\it \noindent 2.1. Notations}
\vglue 0.1cm
Denote by $\BG$ a basic classical Lie superalgebra of rank $r$ $^{13}$.
$\BG=\BH\oplus (\oplus_{\A\in\DE}\BG_{\A})$ is a root space decomposition
of $\BG$.
$\BH$ is the Cartan subalgebra.
The root system $\DE$ of $\BG$ is $\DE^{0}\cup\DE^{1}$, where
$\DE^{0}$ ($\DE^{1}$) is the set of even (odd) roots.
$\DE_{+}=\DE^{0}_{+}\cup\DE^{1}_{+}$ is the set of positive roots
of $\BG$, where $\DE^{0}_{+}$ ($\DE^{1}_{+}$) is the set of even (odd)
positive roots.
$\BG_{0}=\BH\oplus (\oplus_{\A\in\DE^{0}}\BG_{\A})$ is an even subalgebra
of $\BG$. $\BG_{1}=\oplus_{\A\in\DE^{1}}\BG_{\A}$ is an odd subspace.
$\rho_{0}$ ($\rho_{1}$) is half the sum of positive even (odd) roots
and $\rho$ is defined as $\rho_{0}-\rho_{1}$.
$h^{\vee}$ is the dual Coxeter number of $\BG$.
For an affine Lie superalgebra $\HBG$ at level $k$ associated with
$\BG$, we define a constant $\A_{+}=\sqrt{k+h^{\vee}}$.
In the following we shall discuss the Lie superalgebras $A(m|n)$
($h^{\vee}=m-n$) in detail.
\vglue 0.4cm
{\it \noindent 2.2. The Lie superalgebras $A(m|n)$}
\vglue 0.1cm
A Lie superalgebra $sl(n+1|m+1)$  may be represented by matrices
\eq
X=\mat{c|c}{A & a\cr
           \hline
            b & B},
\label{eq:elm}
\en
satisfying $\str X= \tr A-\tr B=0$, where $A$ and $B$ are
$(n+1)\times (n+1)$ and $(m+1)\times (m+1)$ matrices with
grassmann even elements and $a$ and $b$ are
$(n+1)\times (m+1)$ and $(m+1)\times (n+1)$ matrices
with grassmann odd elements.
The commutation relation for two elements
$X_{i}=\mat{c|c}{A_{i} & a_{i}\cr
           \hline
            b_{i} & B_{i}}
$
($i=1,2$) is given by
\eq
[ X_{1}, X_{2} ]
=\mat{c|c}{[ A_{1}, A_{2} ] +a_{1}b_{2}-a_{2}b_{1}&
            a_{1}B_{2}-a_{2}B_{1}+A_{1}a_{2}-A_{2}a_{1} \cr
           \hline
           B_{1}b_{2}-b_{1}B_{2}+b_{1}A_{2}-b_{2}A_{1}&
           [ B_{1}, B_{2} ] +b_{1}a_{2}-b_{2}a_{1} }.
\en
In the case of $n=m$, the identity matrix ${\bf 1}_{2n+2}$
spans an ideal of $sl(n+1|n+1)$ and the Lie superalgebra $A(n|n)$
is defined as $sl(n+1|n+1)/ < {\bf 1}_{2n+2} >$.
Even subalgebras of $A(n|m)$ for $n\neq m$ ($n=m$) are
$A_{n}\oplus A_{m}\oplus u(1)$ ($A_{n}\oplus A_{n}$).
For $A(n|n)$ it is convenient to use a pseudo-representation$^{14}$.
Namely we take a representative of elements of $A(n|n)$ in (\ref{eq:elm})
such that $\tr A =\tr B=0$, and modify the commutation relation like
\eq
 \LB X_{1}, X_{2}  \RB
=[ X_{1}, X_{2} ]-{1\over n+1}\tr (a_{1}b_{2}-a_{2}b_{1}) {\bf 1}_{2n+2} .
\en
In contrast to the simple Lie algebras, there is a variety of
choices of the simple root system of Lie superalgebras, which
correspond to different Dynkin diagrams.
In the case of $m=n-1$ and $n$, we may take the simple roots as purely
odd roots.
For $A(n|n-1)$ they are given by
\eq
\A_{2i-1}=e_{i}-\D_{i}, \quad \A_{2i}=\D_{i}-e_{i+1},
\quad i=1, \ldots, n,
\en
where $e_{i}$ ($i=1, 2,\ldots, n+1$) and $\D_{i}$ ($i=1, \ldots, m+1$)
are orthonormal bases with positive and negative metric.
Similarly the simple roots of $A(n|n)$ are given by
$\A_{1}, \ldots, \A_{2n}, \A_{2n+1}=e_{n+1}-\D_{n+1}$.
The even positive roots of $A(n|m)$ ($m=n,n-1$) are $e_{i}-e_{j}$
($1\leq i < j\leq n+1$) and $\D_{i}-\D_{j}$ ($1\leq i < j\leq m+1$).
The odd positive roots are $e_{i}-\D_{j}$ ($1\leq i \leq j\leq m+1$)
and $\D_{i}-e_{j}$ ($1\leq i < j\leq n+1$).
In the matrix representation of the type (\ref{eq:elm}), we have the
fundamental representation of $\BG$:
\eqn
E_{e_{i}-e_{j}}\!\!\!\!&=&\!\!\!\!
 E_{i,j+1}, \quad
E_{\D_{i}-\D_{j}}= E_{n+1+i, n+2+j}, \CR
E_{e_{i}-\D_{j}}\!\!\!\!&=&\!\!\!\!
E_{i, n+1+j},  \quad
E_{\D_{i}-e_{j}}= E_{n+1+i,j},
\label{eq:rep1}
\enn
for positive roots.
We define $E_{-\A}={}^{t}E_{\A}$ for negative roots $-\A$ ($\A\in\DE_{+}$).
The Cartan elements are defined as $\A\cdot H=[ E_{\A}, E_{-\A} ]$.
Instead of using (\ref{eq:rep1}),
we can take a different representation$^{5,6}$, in which the odd simple
root structure is manifest:
\eq
E_{e_{i}-e_{j}}= E_{2i-1,2j-1}, \quad
E_{\D_{i}-\D_{j}}= E_{2i,2j}, \quad
E_{e_{i}-\D_{j}}= E_{2i-1, 2j}, \quad
E_{\D_{i}-e_{j}}= E_{2i, 2j-1}.
\label{eq:rep2}
\en
Note that in these expressions, $A(n|n-1)$ and $A(n|n)$ have a
structure similar to that of the simple Lie algebras $A_{2n+1}$ and $A_{2n}$,
respectively.

The superalgebra $A(n|m)$ has  rank $n+m+1$, but the rank of $A(n|n)$
reduces to $2n$ due to the existence of the ideal.
Moreover the root vectors are not linearly independent.
In fact, from the relation
$\sum_{i=1}^{n}\LB E_{\A_{2i+1}}, E_{-\A_{2i+1}} \RB =0$,
we have
\eq
\A_{1}+\A_{3}+\cdots+\A_{2n+1}=0.
\label{eq:null}
\en
For both Lie superalgebras $A(n|n-1)$ and $A(n|n)$ half the
sum of positive roots $\rho$ may be shown to be zero.
\vglue 0.4cm
{\it \noindent 2.3. The classical hamiltonian reduction}
\vglue 0.1cm
Let us consider the hamiltonian reduction of the affine Lie superalgebras
$\HBG=A(n|n-1)^{(1)}$ and $A(n|n)^{(1)}$.
On the dual space $\HBG^{*}$ of $\HBG$ one may introduce the
hamiltonian structure which is generated by the coadjoint action or the
gauge transformation:
\eq
\D_{\LM} J(z)=[ \LM(z), \ J(z) ]+\pa\LM(z).
\label{eq:gau}
\en
Let $J_{\A}(z)$ ($\A\in\DE$) and $H^{i}(z)$ ($i=1, \ldots, r$)
be the currents in the canonical basis.
We consider the constraint space ${\cal M}$ in $\HBG^{*}$:
\eq
J_{\A_{i}+\A_{i+1}}(z)=1,  \quad
J_{\A_{i}+\cdots+\A_{j}}(z)=0, \quad \mbox{for $|i-j|>1$}.
\en
In the present case we treat a dynamical system with  second class
constraints$^{15}$ since the fermionic currents $J_{\A_{i}}$ must
satisfy $J_{\A_{i}}(z)J_{\A_{i+1}}(w)\sim 1/(z-w)$.
By introducing auxiliary fermionic fields one can convert these
second class constraints into  first class constraints$^{16}$.
In this extended phase space, one may use the ordinary gauge
fixing procedure.
The first class constrains generate the gauge symmetries in the
extended phase space.
One finds that these gauge symmetries are generated by the
subalgebra $\hat{\bf n}_{-}$, the affine extension of the nilpotent
subalgebra ${\bf n}_{-}$ generated by the negative roots.
The hamiltonian structure on the reduced phase space
${\cal M}/\hat{\bf n}_{-}$
is introduced by projecting the original gauge symmetries
onto the reduced phase space$^{17}$.
This method is also known as  Polyakov's soldering procedure$^{18}$.
One may take  the Drinfeld-Sokolov type gauge for $A(n|n-1)^{(1)}$:
\eqn
J_{DS}(z)&=&\sum_{i=1}^{n+1}U_{n+2-i}(z)E_{n+1,i}
       +\sum_{i=1}^{n}V_{n+1-i}(z)E_{2n+1, n+1+i} \CR
& &+\sum_{i=1}^{n}(G_{n+1-i}(z)E_{n+1, n+1+i}+\bG_{n+1-i}(z)E_{2n+2, i})
+\LM^{A(n|n-1)}
\label{eq:ds1}
\enn
where
\eq
\LM^{A(n|n-1)}=\sum_{i=1}^{n}E_{i,i+1}+\sum_{i=1}^{n-1}E_{n+1+i, n+2+i}
\en
and $V_{1}=U_{1}$.
For $A(n|n)^{(1)}$, we can choose a similar type of gauge fixing.
It is obtained simply by replacing $\LM^{A(n|n-1)}$ by
\eq
\LM^{A(n|n)}=\sum_{i=1}^{n}E_{i,i+1}+\sum_{i=1}^{n}E_{n+1+i, n+2+i}.
\en
\vglue 0.6cm
{\bf \noindent 3. The classical $N=2$ Super $W_{3}$ Algebra}
\vglue 0.4cm
In this section we shall give a non-trivial example of the $N=2$ super
$W$-algebra from the classical hamiltonian reduction of the affine Lie
superalgebra $A(2|1)^{(1)}$.
Let us take the Drinfeld-Sokolov gauge
\eq
J_{DS}(z)=\mat{ccc|cc}{0 & 1 & 0 & 0 & 0 \cr
                    0 & 0 & 1 & 0 & 0 \cr
                U_{3}(z) & U_{2}(z) & U_{1}(z) & G_{2}(z) & G_{1}(z) \cr
                \hline
                   0 & 0 & 0 & 0 & 1 \cr
                \bG_{2}(z) & \bG_{1}(z) & 0 & V_{2}(z) & U_{1}(z)}.
\label{eq:ds}
\en
The gauge transformation (\ref{eq:gau}) has the gauge parameter
\eq
\LM(z)
=\mat{ccc|cc}{
x_{11}(z) & x_{12}(z) & x_{13}(z) & \xi_{11}(z) & \xi_{12}(z) \cr
x_{21}(z) & x_{22}(z) & x_{23}(z) & \xi_{21}(z) & \xi_{22}(z) \cr
x_{31}(z) & x_{32}(z) & x_{33}(z) & \xi_{31}(z) & \xi_{32}(z) \cr
\hline
\eta_{11}(z) & \eta_{12}(z) & \eta_{13}(z) & y_{11}(z) & y_{12}(z) \cr
\eta_{21}(z) & \eta_{22}(z) & \eta_{23}(z) & y_{21}(z) & y_{22}(z) },
\en
where the diagonal elements are parametrized as
$x_{11}=\vep_{1}+2\vep $, $x_{22}=\vep_{2}-\vep_{1}+2\vep $,
$x_{33}=-\vep_{2}+2\vep $ and $y_{11}=\vep_{3}+3\vep $ and
$y_{22}=-\vep_{3}+3\vep $.
By imposing the conditions  that the gauge transformations
preserve the Drinfeld-Sokolov gauge (\ref{eq:ds}), one can reduce
the number of independent gauge parameters and find that all gauge
parameters are expressed in terms of $x_{1}\equiv x_{13}$,
$x_{23}$, $y_{12}$, $\vep$, $\eta_{1}\equiv\eta_{13}$,
$\eta_{2}\equiv\eta_{23}$, $\xi_{1}\equiv\xi_{12}$
and $\xi_{2}\equiv\xi_{22}$.

Note that we may regard the bosonic part of this system as the coupled
one of $gl(3)$ and $gl(2)$ $W$ algebras which share the same $u(1)$
current $U_{1}$.
But if one requires the $N=2$ superconformal symmetry
which is generated by $\xi_{2}$ and $\eta_{2}$, one finds that
the $N=2$ supermultiplet of the $u(1)$ current $U_{1}$ is
$(U_{1}, G_{1}, \bG_{1}, U_{2}-V_{2})$.
Hence $T\equiv U_{2}-V_{2}$ becomes the energy-momentum tensor,
and $V_{2}$ turns out to be a spin two field.
Corresponding to this change of physical variables, we redefine the gauge
parameters  as $x=x_{23}+y_{12}$ and $y=y_{12}$ instead of $x_{23}$
and $y_{12}$.
One may add suitable differential polynomials of fields
in order to get well defined primary fields.
This can be done by replacing a gauge parameter $\vep$ by
\eq
\vep+{1\over 6}\{ U_{1}x+3\pa x -3U_{1}y -3 \pa y
+2U_{2}x_{1}-2\pa(U_{1}x_{1})-2\pa^2 x_{1}
+3G_{1}\eta_{1}-2\bG_{1}\xi_{1}\} .
\en
After this change of variables, one gets the gauge transformations on the
reduced phase space of the form:
\eq
\D X_{i} =\cD_{i j} Y_{j},
\label{eq:gtrrd}
\en
where $X=^{t}( U_{1},T,V_{2}, U_{3}, G_{1}, G_{2},\bG_{1}, \bG_{2} )$,
$Y=^{t}( \vep, x, y, x_{1}, \xi_{1}, \xi_{2}, \eta_{1}, \eta_{2})$
and $\cD$ is a $8\times 8$ matrix valued differential operator which
satisfies $\cD_{j i}=\cD_{i j}^{*}$, where for a differential operator
$\sum_{i}a_{i}(z)\pa^{i}$, a formal adjoint $*$ is defined as
$(\sum_{i}a_{i}(z)\pa^{i})^{*}=\sum_{i}(-\pa)^{i}a_{i}(z)$.

Let us introduce the Poisson bracket structure by expressing $\D_{\LM}$ as
\eqn
\D_{\LM}&=&\int {d z\over 2\pi\I}
         \str \{ \mat{ccc|cc}{
2\vep & 0 & x_{1} & 0 & \xi_{1} \cr
0 & 2\vep & x-y & 0 & \xi_{2} \cr
0 & 0 & 2\vep & 0 & 0 \cr
\hline
0 & 0 & \eta_{1} & 3\vep & y \cr
0 & 0 & \eta_{2} & 0 & 3\vep }
\mat{ccc|cc}{0 & 0 & 0 & 0 & 0 \cr
                    0 & 0 & 0 & 0 & 0 \cr
                U_{3} & T+V_{2} & U_{1} & G_{2} & G_{1} \cr
                \hline
                   0 & 0 & 0 & 0 & 0 \cr
                \bG_{2} & \bG_{1} & 0 & V_{2} & U_{1}} \}  \CR
&=& \int {d z\over 2\pi\I}
    ( U_{3}x_{1}+T x-y V_{2}-\vep U_{1}
     +\xi_{1}\bG_{2}+\xi_{2}\bG_{1}
     -\eta_{1}G_{2}-\eta_{2}G_{1} ).
\label{eq:gauge}
\enn
{}From Eqs. (\ref{eq:gtrrd}) and (\ref{eq:gauge}), we get the Poisson
bracket structure on the reduced phase space in the form of
operator product expansions.
First we write down the $N=2$ superconformal algebra:
\eqn
T(z)T(w)&\sim & \zw{2}{T(z)+T(w)}, \quad
T(z)U_{1}(w) \sim \zw{3}{6}+\zw{2}{U_{1}(z)}, \CR
\bG_{1}(z)T(w)&\sim &\zw{2}{\bG_{1}(z)+\bG_{1}(w)}, \quad
G_{1}(z)T(w)\sim \zw{2}{G_{1}(w)},  \CR
U_{1}(z)U_{1}(w)&\sim & \zw{2}{-6}, \CR
U_{1}(z)G_{1}(w)&\sim & \zw{}{G_{1}(w)}, \quad
U_{1}(z)\bG_{1}(w)\sim  \zw{}{-\bG_{1}(w)}, \CR
G_{1}(z)\bG_{1}(w)&\sim &\zw{3}{-6}
                    +\zw{2}{U_{1}(w)}
                    +\zw{}{T(w)}.
\enn
Note that in this expression $T(z)$ is  the twisted $N=2$
energy-momentum tensor.
Second the $N=2$ supermultiplet structures for
$(V_{2},G_{2}, \bG_{2}, U_{3})$ are given by
\eqn
T(z)V_{2}(w)&\sim & \zw{4}{6}+\zw{3}{-U_{1}(w)}
+\zw{2}{V_{2}(w)+V_{2}(z)}, \CR
T(z)U_{3}(w)&\sim &\zw{2}{U_{3}(z)+2U_{3}(w)}, \CR
T(z)\bG_{2}(w)&\sim &\zw{2}{2\bG_{2}(w)+\bG_{2}(z)}, \quad
T(z)G_{2}(w)\sim \zw{3}{-2G_{1}(w)}+\zw{2}{G_{2}(w)+G_{2}(z)}, \CR
U_{1}(z)V_{2}(w)&\sim & \zw{3}{-6}+\zw{2}{3U_{1}(w)}, \CR
U_{1}(z)U_{3}(w)&\sim & \zw{4}{-12}+\zw{3}{4U_{1}(w)}
                    +\zw{2}{2(T+V_{2})(w)}, \CR
U_{1}(z)G_{2}(w)&\sim & \zw{2}{3 G_{1}(w)}+\zw{}{G_{2}(w)}, \quad
U_{1}(z)\bG_{2}(w)\sim  \zw{2}{2 \bG_{1}(w)}+\zw{}{-\bG_{2}(w)}, \CR
\bG_{1}(z)V_{2}(w)&\sim &\zw{2}{\bG_{1}(w)}
               +\zw{}{\bG_{2}(w)}, \quad
G_{1}(z)V_{2}(w)\sim \zw{}{-G_{2}(w)}, \CR
\bG_{1}(z)U_{3}(w)&\sim &\zw{2}{2\bG_{2}(w)+\bG_{2}(z)} , \quad
G_{1}(z)U_{3}(w)\sim \zw{3}{4G_{1}(w)}+\zw{2}{2G_{2}(w)}, \CR
\bG_{2}(z)G_{1}(w)&\sim & \zw{4}{-12}
                    +\zw{3}{-4U_{1}(z)}
                    +\zw{2}{2V_{2}(z)}
                    +\zw{}{-U_{3}(w)}, \CR
\bG_{1}(z)G_{2}(w)&\sim &\zw{4}{6}
                    +\zw{3}{-2U_{1}(w)}
                    +\zw{2}{V_{2}(w)+V_{2}(z)+T(w)}
                    +\zw{}{-U_{3}(w)}.
\enn
Finally the remaining nontrivial operator product expansions take the forms:
\eqn
V_{2}(z)V_{2}(w)\!\!\!\! &\sim &\!\!\!\!
\zw{4}{-12}+\zw{3}{4[U_{1}(w)-U_{1}(z)]}
+\zw{2}{V_{2}(z)+V_{2}(w)}+\zw{}{2U_{1}(z)U_{1}(w)}, \CR
U_{3}(z)V_{2}(w)\!\!\!\! &\sim &\!\!\!\!
      \zw{5}{-24}+\zw{4}{6[U_{1}(w)-U_{1}(z)]}
      +\zw{3}{2[ U_{2}(z)+U_{1}(z)U_{1}(w) ]}     \CR
        & &\!\!\!\!   +\zw{2}{-U_{1}(w)U_{2}(z)+\bG_{1}(w)G_{1}(z)}
              +\zw{}{[\bG_{2}G_{1}-\bG_{1}G_{2}](w)}, \CR
U_{3}(z)U_{3}(w)\!\!\!\!&\sim &\!\!\!\!
              \zw{3}{2[U_{3}(z)-U_{3}(w)]} \CR
        & &      +\zw{2}{-U_{3}(w)U_{1}(z)-U_{1}(w)U_{3}(z)
                      +\bG_{2}(w)G_{1}(z)-G_{1}(w)\bG_{2}(z)}, \CR
\bG_{2}(z)V_{2}(w)\!\!\!\!&\sim &\!\!\!\!
                \zw{3}{2[\bG_{1}(z)-\bG_{1}(w)]} \CR
          & &     +\zw{2}{-\bG_{2}(w)-\bG_{1}(w)U_{1}(z)-U_{1}(w)G_{1}(z)}
               +\zw{}{[\bG_{1}V_{2}-\bG_{2}U_{1}](w)}, \CR
G_{2}(z)V_{2}(w)\!\!\!\!&\sim &\!\!\!\!
                \zw{3}{4G_{1}(z)}
               +\zw{2}{-2U_{1}(w)G_{1}(z)-G_{2}(z)}
               +\zw{}{[U_{1}G_{2}-V_{2}G_{1}](w)}, \CR
\bG_{2}(z)U_{3}(w)\!\!\!\!&\sim &\!\!\!\!
                \zw{3}{2[\bG_{2}(z)-\bG_{2}(w)]}
               -\zw{2}{U_{1}(w)\bG_{2}(z)+\bG_{2}(w)U_{1}(z)}
               +\zw{}{[U_{3}\bG_{1}-T\bG_{2}](w)}, \CR
G_{2}(z)U_{3}(w)\!\!\!\!&\sim &\!\!\!\!
               \zw{4}{6[G_{1}(z)-G_{1}(w)]}
               +\zw{3}{-2G_{2}(w)-2G_{1}(w)U_{1}(z)-2U_{1}(w)G_{1}(z)} \CR
 & &\!\!\!\!\!  +\zw{2}{-G_{2}(w)U_{1}(z)-G_{1}(w)T(z)-U_{2}(w)G_{1}(z)}
               +\zw{}{-[G_{2}T+G_{1}U_{3}](w)}, \CR
\bG_{2}(z)G_{2}(w)\!\!\!\!&\sim& \!\!\!\!
\zw{5}{-24}+\zw{4}{6[U_{1}(w)-U_{1}(z)]}
+\zw{3}{2[U_{1}(w)U_{1}(z)+V_{2}(z)+T(w)]}  \CR
& & \!\!\!\!\!\!\!\!\!\!\!\!\!\!\!\!\!\!\!\!
+\zw{2}{U_{3}(w)+T(w)U_{1}(z)-U_{1}(w)V_{2}(z)+G_{1}(w)\bG_{1}(z)}
+\zw{}{[U_{3}U_{1}-V_{2}T](w)},  \CR
G_{2}(z)G_{2}(w)\!\!\!\! &\sim & \!\!\!\!
 \zw{}{[2G_{1}\pa G_{1}+2G_{2}G_{1}](w)}, \quad
 \bG_{2}(z)\bG_{2}(w)\sim  \zw{}{2\bG_{1}\bG_{2}(w)} .
\enn
The present Poisson bracket structure is the same as that obtained from
the super Gel'fand-Dickii algebra$^{9,10}$ but different from the results
in ref. 8 due to the different choice of the gauges.
\vglue 0.6cm
{\bf \noindent 4. The quantum hamiltonian reduction and
$N=2$ $CP_{n}$ coset models}
\vglue 0.4cm
So far we have discussed the classical hamiltonian reduction.
In the quantum case, we use the BRST gauge fixing procedure by
introducing ghost systems for the constraints$^{3,2}$.
In order to impose the constraints at the quantum level, we
must improve the energy-momentum tensor $T_{WZNW}$ of the
Wess-Zumino-Novikov-Witten model corresponding to the affine Lie
superalgebra $\HBG$ by the Cartan currents $H^{i}(z)$,
\eq
T_{improved}(z)=T_{WZNW}(z)+\mu\cdot\pa H(z).
\en
After this improvement, the conformal dimensions of the currents
$J_{\A}(z)$ become $1-\mu\cdot\A$.

First we discuss the $A(n|n-1)^{(1)}$ case.
The improvement vector $\mu=\mu_{A(n|n-1)}$ is defined by the conditions:
\eq
\A_{i}\cdot\mu_{A(n|n-1)}=\half, \quad i=1,\ldots, 2n.
\label{eq:imp1}
\en
After this improvement, the conformal dimensions of the currents
for the odd simple roots $J_{\A_{i}}(z)$ becomes $\half$ and zero for the
the currents for the simple roots $\A_{i}+\A_{i+1}$ of the even subalgebras.
Hence we may introduce the \lq\lq diagonal" gauge:
\eq
J_{diag}(z)= \pa \vp(z)\cdot H
             + \sum_{i=1}^{n}
    ( \A_{2i-1}\cdot\chi(z)E_{\A_{2i-1}}+\A_{2i}\cdot\chi(z)E_{\A_{2i}})
             +\LM^{A(n|n-1)},
\label{eq:dia1}
\en
where $\chi^{i}(z)$ ($i=1,\ldots, 2n$) are $2n$ real fermions,
satisfying  $\chi^{i}(z)\chi^{j}(w)\sim\D_{i j}/(z-w)$.
The improvement vector $\mu_{A(n|n-1)}$ is uniquely determined
by (\ref{eq:imp1}) and is expressed as
\eq
\mu_{A(n|n-1)}
=\half\sum_{i=1}^{n}[ (n+1-i)\A_{2i-1}+i\A_{2i} ]
=\half\sum_{i=1}^{2n}\lm_{i},
\en
where $\lm_{i}$ are fundamental weights of $A(n|n-1)$ satisfying
$\lm_{i}\cdot\A_{j}=\D_{i j}$:
\eq
\lm_{2i}=\A_{1}+\A_{3}+\cdots+\A_{2i-1}, \quad
\lm_{2i-1}=\A_{2 i}+\A_{2i+2} +\cdots+\A_{2n},
\en
for $i=1, \ldots, n$.
By using the Wakimoto realization of affine Lie superalgebras,
one finds that the total energy-momentum tensor
$T_{total}=T_{improved}+T_{ghosts}+T_{\chi}$
is BRST-equivalent to that of the $N=2$ $CP_{n}$ coset model$^{5,7}$:
\eq
T_{total}(z)=T_{CP_{n}}(z)+\{ Q_{BRST}, * \},
\en
where $T_{\chi}=-\half\sum_{i=1}^{2n}\chi^{i}\pa\chi^{i}$ and
\eq
T_{CP_{n}}(z)=-\half (\pa\vp)^{2}
-\I\A_{+}\mu_{A(n|n-1)}\cdot\pa^{2}\vp +T_{\chi}.
\en
The remaining $N=2$ generators are expressed as
\eqn
U_{1}(z)&=&\sum_{i=1}^{n}\lm_{2i}\cdot\chi \A_{2i}\cdot\chi
      +\I \A_{+} \nu \cdot \pa \vp, \CR
G_{1}(z)&=&\sum_{j=1}^{n}
     (\I\A_{2j}\cdot \pa \vp \lm_{2j} \cdot \chi
      - \A_{+}\lm_{2j}\cdot \pa\chi), \CR
\bG_{1}(z)&=&\sum_{j=1}^{n}
        (\I\A_{2j-1}\cdot \pa \vp \lm_{2j-1} \cdot \chi
         -\A_{+}\lm_{2j-1}\cdot \pa\chi),
\enn
where $\nu=\sum_{i=1}^{n}(\lm_{2i}-\lm_{2i-1})$.
The other $W$ currents can be obtained by the quantum Miura transformation,
which connects the diagonal gauge (\ref{eq:dia1}) with the
Drinfeld-Sokolov gauge (\ref{eq:ds1}).
The free field representation of the $N=2$ $CP_{n}$ models has been
studied in ref. 6 in detail.

Now we proceed to the affine Lie superalgebra $A(n|n)^{(1)}$.
In this case one cannot impose the spin $\half$ constraints
(\ref{eq:imp1}) for
the simple roots $\A_{i}$ ($i=1,\ldots, 2n+1$) due to the relation
(\ref{eq:null}).
Instead we should require the conditions:
\eq
\A_{2i-1}\cdot\mu=0, (i=1, \ldots, n+1),
\quad \A_{2j}\cdot\mu=1, (j=1,\ldots, n),
\en
that are consistent with (\ref{eq:null}).
This means that the conformal weights are 1 for $J_{\A_{2i-1}}$ but
0 for $J_{\A_{2i}}$ after the improvement.
The vector $\mu$ is determined uniquely
up to $\sum_{i=1}^{n+1}\A_{2i-1}(=0)$:
\eq
\mu=\mu_{A(n|n)}=\sum_{i=1}^{n+1}(n+1-i)\A_{2i-1}.
\en
The diagonal gauge for $A(n|n)^{(1)}$ is
\eq
J_{diag}(z)= \pa \vp(z)\cdot H
             + \sum_{i=1}^{n+1}
              \eta_{2i-1}(z)E_{\A_{2i-1}}
             +\sum_{i=1}^{n} \xi_{2i}(z)E_{\A_{2i}}
             +\LM^{A(n|n)},
\en
where $\eta_{2i-1}=\eta_{2i}-\eta_{2i-2}$
($\eta_{2n+2}=\eta_{0}\equiv 0$) and
($\eta_{2i}, \xi_{2i}$) are fermionic ghosts with conformal weight $(1,0)$.
The energy-momentum tensor of the reduced theory  becomes
\eq
\tilde{T}_{CP_{n}}(z)=-\half (\pa\vp)^{2}
-\I\A_{+}\mu_{A(n|n)}\cdot\pa^{2}\vp
-\sum_{i=1}^{n}\eta_{2i}\pa\xi_{2i},
\en
which is equal to $T_{CP_{n}}+\half \pa U_{1}$.
This is nothing but the energy-momentum tensor of
the twisted$^{19}$ $N=2$ $CP_{n}$ model.
Hence the conformal field theory corresponding to the
affine Lie superalgebra $A(n|n)^{(1)}$ should be regarded as a topological
conformal field theory rather than an $N=2$ superconformal field theory.
This implies that the $A(n|n)$ Toda field theory has the {\em twisted}
$N=2$ super $W$-algebra symmetry instead of the $N=2$ superconformal
symmetry.
Recently Evans and Hollowood$^{20}$ also pointed out that the $A(n|n)$
Toda field theory does not have $N=2$ superconformal symmetry.

In this sense the $A(n|n)^{(1)}$ affine Toda field theory
can be also regarded as a topological field theory rather than an $N=2$
theory.
It is a quite interesting problem to study this affine Toda field theory
as a topological field theory since this model gives a different
class of topological solvable models which are not classified by the
$N=2$ Landau-Ginzburg type models.
\vglue 0.6cm
{\bf \noindent Acknowledgements}
\vglue 0.4cm
The author would like to thank Jens Lyng Petersen for helpful comments.
This work was supported in part by EEC contract SC1 394 EDB.
\vglue 0.6cm
{\bf \noindent References}
\vglue 0.4cm
\newcommand{\NP}{{\it Nucl.~Phys.}}
\newcommand{\PL}{{\it Phys.~Lett.}}
\newcommand{\CMP}{{\it Commun.~Math.~Phys.}}
\newcommand{\MPL}{{\it Mod.~Phys.~Lett.}}
\newcommand{\IJMP}{{\it Int.~J.~Mod.~Phys.}}

\end{document}